\documentclass[twocolumn,showpacs,superscriptaddress,aps,prl,amsfonts,reprint,preprintnumbers,amsmath,amssymb]{revtex4}
\usepackage{wrapfig,subfig,graphicx,ulem}
\usepackage{xcolor}
\usepackage{dcolumn} 
\usepackage{bm} 
\usepackage{natbib}

\newcommand\tb{\textbf}

\newcommand\p{\partial}
\newcommand\beq{\begin{equation}}
\newcommand\eeq{\end{equation}}

\begin{document}
\title{Micron-Scale Mapping of Megagauss Magnetic Fields in Petawatt Laser-Solid Interactions}
\author{Gourab Chatterjee}
\author{Prashant Kumar Singh}
\affiliation{Tata Institute of Fundamental Research, Homi Bhabha Road, Mumbai 400005, India}
\author{A. P. L. Robinson}
\affiliation{Central Laser Facility, STFC Rutherford Appleton Laboratory, Chilton, Didcot OX11 0QX, UK}
\author{N. Booth}
\affiliation{Central Laser Facility, STFC Rutherford Appleton Laboratory, Chilton, Didcot OX11 0QX, UK}
\author{O. Culfa}
\author{R. J. Dance}
\affiliation{York Plasma Institute, University of York, Heslington, York YO10 5DQ, UK}
\author{L.  A. Gizzi}
\affiliation{Intense Laser Irradiation Laboratory (ILIL), INO-CNR, Pisa, Italy}
\author{R. J. Gray}
\affiliation{SUPA, Dept. of Physics, University of Strathclyde, Glasgow G4 0NG, UK}
\author{J. S. Green}
\affiliation{Central Laser Facility, STFC Rutherford Appleton Laboratory, Chilton, Didcot OX11 0QX, UK}
\author{P. Koester}
\affiliation{Intense Laser Irradiation Laboratory (ILIL), INO-CNR, Pisa, Italy}
\author{G. Ravindra Kumar}
\affiliation{Tata Institute of Fundamental Research, Homi Bhabha Road, Mumbai 400005, India}
\author{L. Labate}
\affiliation{Intense Laser Irradiation Laboratory (ILIL), INO-CNR, Pisa, Italy}
\author{Amit D. Lad}
\affiliation{Tata Institute of Fundamental Research, Homi Bhabha Road, Mumbai 400005, India}
\author{K. L. Lancaster}
\affiliation{Central Laser Facility, STFC Rutherford Appleton Laboratory, Chilton, Didcot OX11 0QX, UK}
\author{J. Pasley}
\affiliation{Central Laser Facility, STFC Rutherford Appleton Laboratory, Chilton, Didcot OX11 0QX, UK}
\affiliation{York Plasma Institute, University of York, Heslington, York YO10 5DQ, UK}
\author{N. C. Woolsey}
\affiliation{York Plasma Institute, University of York, Heslington, York YO10 5DQ, UK}
\author{P. P. Rajeev}
\email{Rajeev.Pattathil@stfc.ac.uk; pprajeev@tifrh.res.in}
\affiliation{Central Laser Facility, STFC Rutherford Appleton Laboratory, Chilton, Didcot OX11 0QX, UK}
\affiliation{TIFR Centre for Interdisciplinary Sciences, Tata Institute of Fundamental Research, Hyderabad, India}

\begin{abstract}
\noindent{We report spatially and temporally resolved measurements of magnetic fields generated by petawatt laser-solid interactions with high spatial resolution, using optical polarimetry. The polarimetric measurements map the megagauss magnetic field profiles generated by the fast electron currents at the target rear. The magnetic fields at the rear of a 50 $\mu$m thick aluminum target exhibit distinct and unambiguous signatures of electron beam filamentation.  These results are corroborated by hybrid simulations.}         
\end{abstract}

\pacs{52.38.Fz, 52.25.Fi, 52.70.Kz}

\maketitle

The interaction of an intense laser pulse with a solid generates mega-amperes of relativistic `fast' electron bunches, which produce the largest terrestrial magnetic fields with magnitudes approaching gigagauss levels \cite{SudanPRL1993,TatarakisNature2002, WagnerPRE2004}. These magnetic fields are pivotal in determining the propagation of the fast electrons that generate them \cite{AlfvenPR1939,WeibelPRL1959}, leading to a complex interplay between the fast electrons and the magnetic fields. Fast electrons are of critical importance to a number of potential applications, including the development of novel x-ray sources \cite{MurnaneScience1991}  and alternate particle acceleration schemes \cite{SnavelyPRL2000}, simulation of astrophysical conditions in the laboratory \cite{RemingtonRMP2006} and the fast ignition variant of inertial fusion \cite{TabakPOP1994}. Measuring magnetic fields can shed light on the fast electron distribution inside solids \cite{WilksPRL1992, SarriPRL2012}.  Since the interplay between the fast electrons and the self-generated magnetic fields is inherently transient and confined to micron-scales, diagnostics with high spatio-temporal resolution are essential for a better understanding of the fast electron transport process. 

In this Letter, we present measurements of the megagauss magnetic fields generated at the rear of solid targets irradiated with a petawatt laser, using optical Cotton-Mouton polarimetry \cite{SandhuPRL2002, KahalyPOP2009, MondalPNAS2012}. This technique offers an unprecedented spatio-temporal resolution in the mapping of the self-generated magnetic fields, yielding new insights into the principal characteristics of fast electron transport through solids. For instance, this technique enables us to temporally resolve, for the first time,  signatures of  micron-level filamentary instabilities in the fast electron transport through a metallic (aluminum) target at petawatt laser irradiances. Three-dimensional (3D) hybrid simulations clearly exhibit these filamentary features, attributed to the fast electron transport in metals being subjected to resistive filamentation at sufficiently high temperatures.

\begin{figure}[b]
\centering\includegraphics[width=\columnwidth]{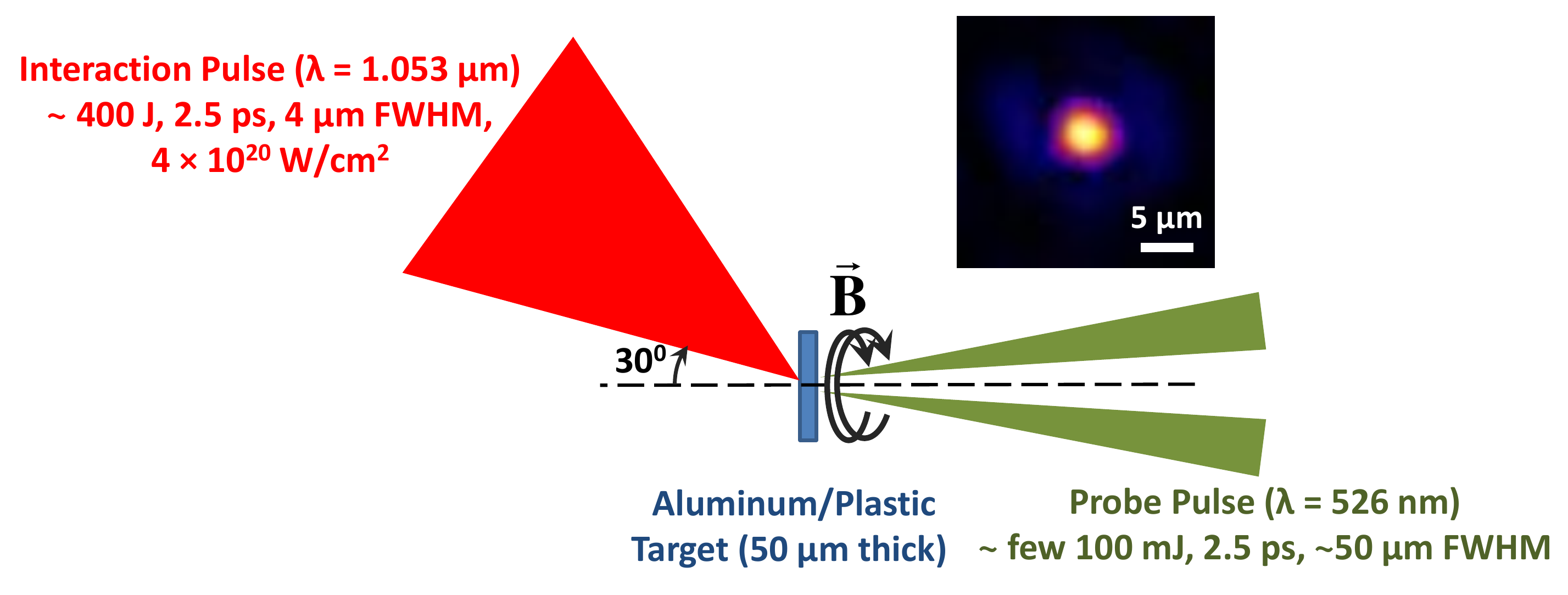}
\caption{\footnotesize{Schematic of the experimental setup. Inset shows the typical transverse profile of the focal spot of the interaction laser pulse on the target.}} 
\end{figure}

Various innovative techniques have been employed to probe magnetic fields generated in the plasma at the target front.   For instance, the $X$-wave cutoff of laser-generated harmonics \cite{TatarakisNature2002, WagnerPRE2004} can be used to measure magnetic fields at the critical surface on the front side. Faraday rotation of an external probe \cite{StamperPRL1975,BorghesiPRL1998} can provide temporal snapshots of magnetic fields in an underdense plasma,  integrated along the transverse density profile. Proton deflectometry can also provide time-resolved magnetic field measurements \cite{WillingalePRL2010, SarriPRL2012}, but integrated along the thickness of the target.

In order to spatially resolve fast electron transport in longitudinal as well as transverse directions, one needs to probe the magnetic fields in the plasma at the target \textit{rear} surface \cite{DaviesPRE1999}, resolved along the transverse plane. These magnetic field structures are determined by the fast electron transport pattern in the bulk target. Thus, detailed measurements of these fields allow one to make detailed inferences about transport in the bulk target as well as detailed comparisons with numerical models. To understand the dynamics of fast electron transport, a complete spatio-temporal characterisation of the magnetic fields is therefore  required.  However, there are limited experimental techniques available to this effect.  To achieve a complete spatio-temporal mapping of the magnetic fields, we resort to a time-delayed optical probe reflected off its critical surface at the target rear. 

The experiment was performed at the Rutherford Appleton Laboratory using the Vulcan Petawatt laser, delivering more than 400 J on target at a central wavelength of 1.053 $\mu$m over a pulse duration of 2.5 ps. A schematic of the experimental setup is shown in Fig. 1. The $p$-polarized interaction laser pulse was focused on the target by an $f/3$ off-axis parabolic mirror at an angle of incidence of 30$^0$. Measurements at low intensities estimated the focal spot to be 4 $\mu$m (FWHM), containing about 30\% of the laser energy in the focal volume, resulting in a peak intensity of $4\times10^{20}$ W/cm$^2$.  A linearly-polarized, time-delayed and frequency-doubled ($\lambda=526$ nm) probe pulse, extracted from the main interaction pulse, was  suitably attenuated to low intensities and focused  to a 50 $\mu$m diameter spot on the target rear at near-normal incidence. The magnetic fields induce a birefringence in the plasma at the target rear, resulting in a change in the polarization state of the incident probe, which was inferred from standard polarimetric measurements of the Stokes' parameters of the reflected probe \cite{Hutchinsonbook, SegrePPCF1999, ellipticity}. An optical streak camera was employed to synchronize the probe pulse with the main interaction pulse, the synchronization being limited by the pulsewidth of 2.5 ps. All temporal delays between the probe and the main interaction pulse are peak-to-peak measurements. Charge-coupled-device (CCD) cameras with interference filters were used to monitor the spatial profile of the magnetic fields. The polarimetric measurements indicated an induced ellipticity in the probe due to the azimuthal nature of the self-generated magnetic fields, consistent with previous experiments and simulations. The Faraday rotation of the normally incident probe due to any axial component of the magnetic field was found to be below the threshold of detection. 1D radiation hydrodynamics simulations using the HYADES code provide the scale-length of the plasma density profile at the target rear, assuming that the target rear is volume heated to a temperature consistent with that observed for similar targets in previous experiments under similar conditions, as for example in reference \cite{LancasterPOP2009}. Further details on the polarimetric setup and the evaluation of magnetic fields can be found in our previous works \cite{SandhuPRL2002, KahalyPOP2009, MondalPNAS2012}. 

\begin{figure}[b]
\includegraphics[width=0.6\columnwidth]{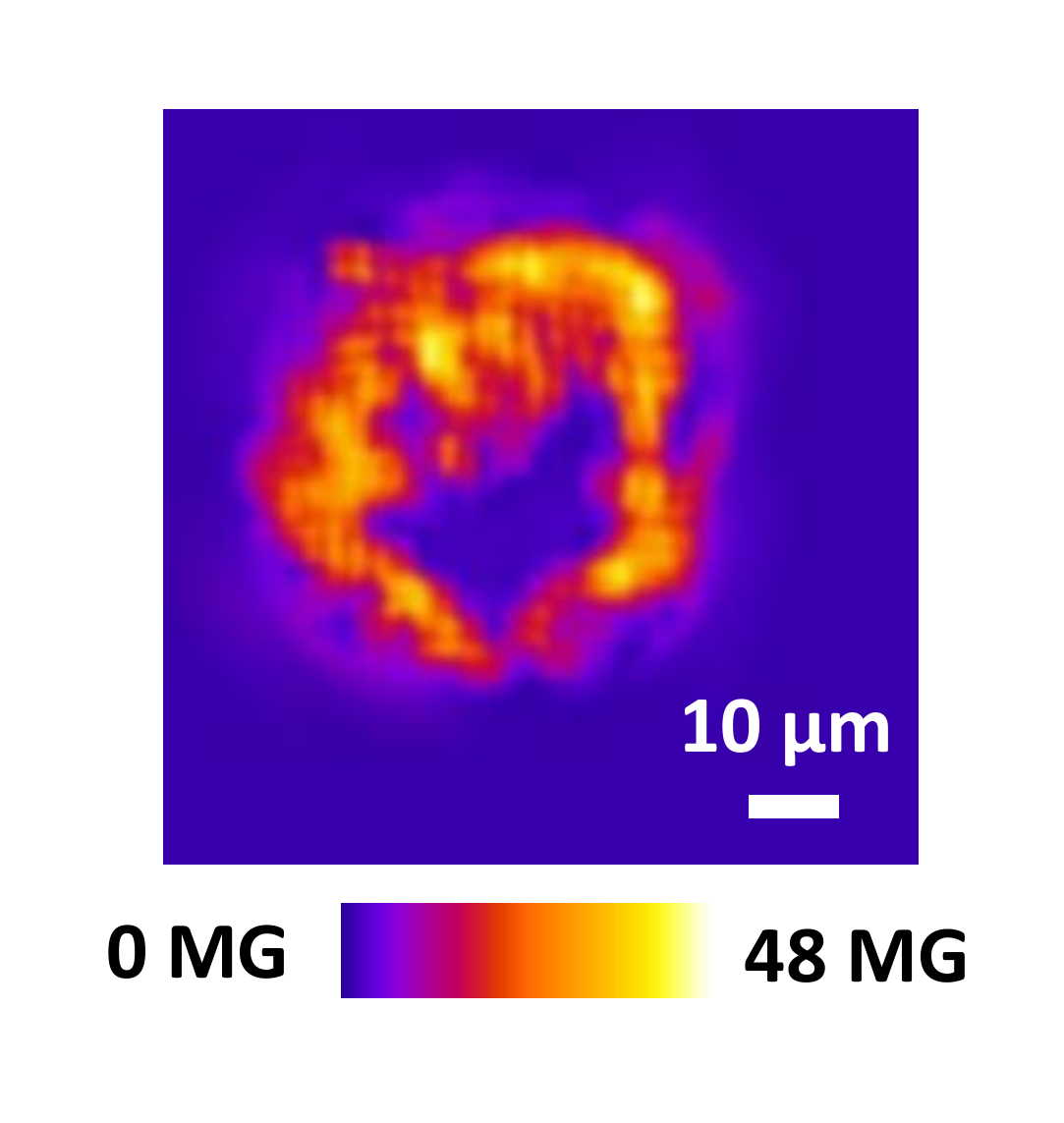}
\caption{\footnotesize{The magnetic field profile at the rear of a 1 mm $\times$ 1 mm, 50 $\mu$m thick plastic target (with a 2 $\mu$m aluminum coating at the rear surface for a specular reflection of the optical probe), at a temporal delay of 5 ps after the interaction pulse. The annular profile of the magnetic field is indicative of a beam-like distribution of the fast electrons exiting the target rear surface. The spatial resolution of the optical imaging setup is $\sim 10$ $\mu$m and the error in the estimation of the peak magnetic field is $\pm 5$ MG.}}
\end{figure}

To exemplify the efficacy of the aforesaid polarimetric technique, we present the magnetic field profile (Fig. 2) at the rear of a 50 $\mu$m thick plastic (mylar) target, 5 ps after the interaction pulse. The magnetic field reaches local peak values of $\sim$ 50 MG,  although the value of the magnetic field spatially averaged over the 50 $\mu$m diameter probe focal spot is about 3 MG. The most significant feature in the profile, however, is the annular distribution of the magnetic field with a central hollow, as seen in previous simulations \cite{HondaPOP2000, PukhovPRL2001, SentokuPRE2002}. Such an annular magnetic field profile at the target rear is indicative of a beam-like distribution of the fast electrons exiting the target \cite{PukhovPRL2001}. In addition, despite the limited spatial resolution of this measurement ($\sim 10$ $\mu$m), the magnetic field profile exhibits the onset of filamentation at the periphery, indicating that the fast electron beam is beginning to fragment inside the dielectric \cite{GremilletPOP2002}.

The current understanding is that fast electron propagation through metals is less prone to filamentation, compared to materials that are dielectric or insulating at room temperature. This has been supported by several measurements of inferred electron beam profiles, employing proton radiography \cite{FuchsPRL2003} or optical emission \cite{ManclossiPRL2006}. A smooth beam profile was observed in metallic targets up to tens of microns in thickness, whilst the profile had distinct filamentary features when the fast electrons traversed through dielectric targets. However, it is not clear that this material dependence should be universal, as the details of the interaction should also depend on the properties of the fast electron beams and the local target conditions, which, in turn, are dependent on laser parameters like intensity, pulsewidth, contrast etc. For instance, recent experiments \cite{Scott_PRL_2012} as well as simulations \cite{OvchinnikovPRL2013} indicate that the divergence of the fast electron beam seems to be clearly dependent on the preplasma conditions in metallic targets. Depending on the temperature-dependent resistivity \cite{BellPPCF1997}, the fast electron beam profile inside metals can also be filamented, as shown in recent simulations \cite{SarriPRL2012, SentokuPRL2011} and time-integrated measurements \cite{StormPRL2009}. Since temperature and hence resistivity are transient in these experimental conditions, it is essential to have a diagnostic with sufficient spatial and temporal resolution to resolve these filamentary structures. Our diagnostic has sufficient resolution to observe even micron-scale filamentation with picosecond-resolution.  

We now present results from magnetic field measurements at the rear of a 50 $\mu$m thick aluminum target irradiated under similar conditions with micron-scale resolution, as shown in Fig. 3a.  Firstly, unlike plastic, the magnetic field here is clearly not annular, suggesting that the fast electrons are not ``beam-like" as they traverse through aluminum.  Secondly, there are distinct micron-scale filamentary structures in the magnetic field distribution as shown in the magnified view of the magnetic field profile (Fig. 3b). Although the above measurement is at a temporal delay of 10 ps, similar measurements even a few picoseconds after the main interaction pulse exhibited distinct signatures of filamentation, whilst measurements made before the arrival of the main interaction pulse showed a uniformly null magnetic field profile, indistinguishable from the background (see the Supplemental Material \cite{Supplemental}). Figure 3 shows the filaments reaching local peak magnetic field values over 100 MG, interspersed with regions of near-zero magnetic fields, indicating a heavily fragmented electron distribution. The curl of the magnetic field profile ($\nabla\times\tb B$) yields the spatial profile for the current density, which exhibits peak values of $\sim 3\times 10^{11}$ A/cm$^2$. Integrating over a typical filament size yields a maximum net forward current of $\sim 30$ kA, which is below the Alfven limit \cite{AlfvenPR1939}.

\begin{figure}[t]
\includegraphics[width=\columnwidth]{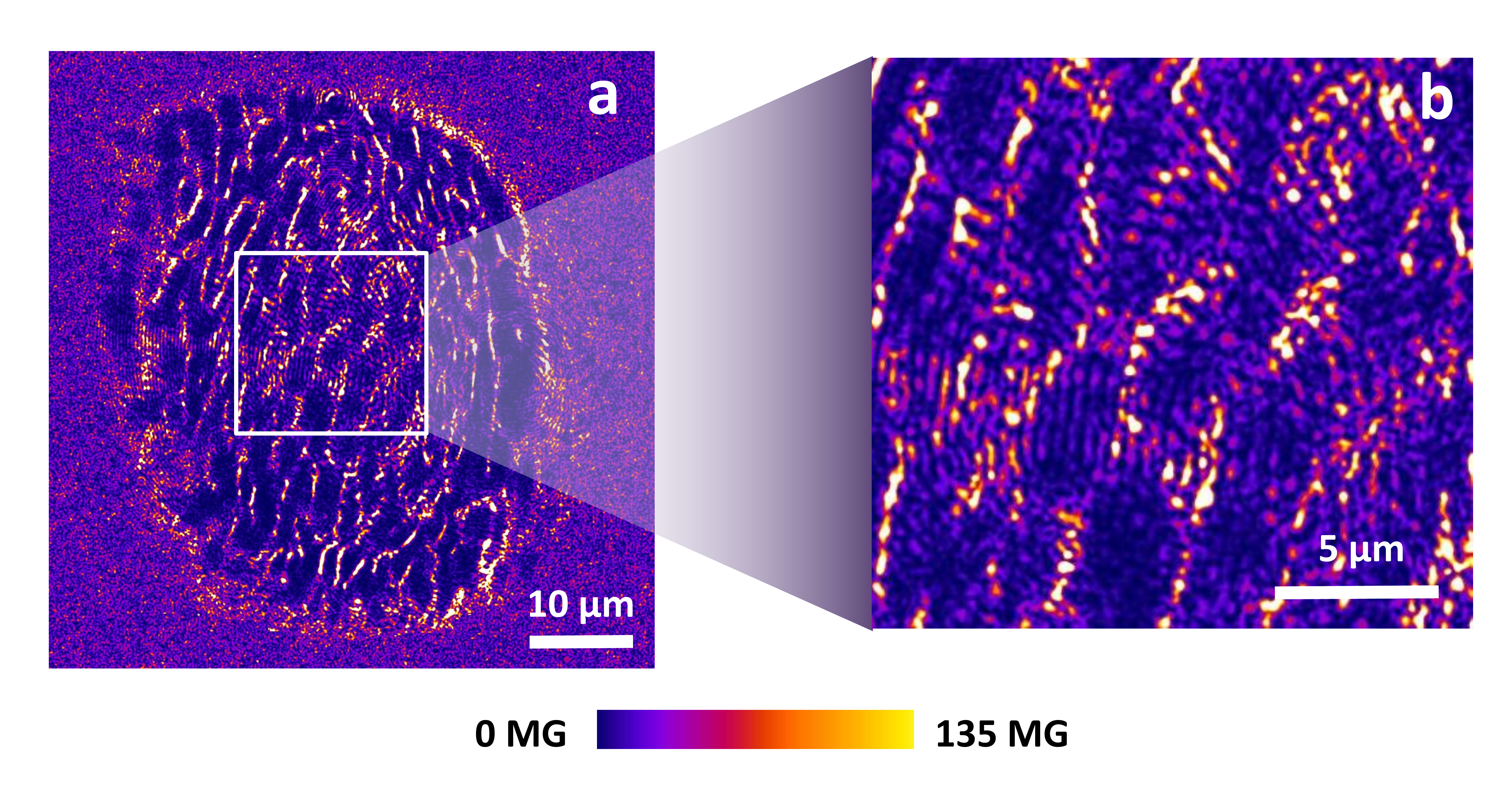}
\caption{\footnotesize{(a) The magnetic field profile  at the rear of a 1mm $\times$ 1 mm, 50 $\mu$m thick aluminum target, 10 ps after the interaction pulse. (b) Magnified view of a section of the magnetic field profile, highlighting the filamentary structures. The spatial resolution of the optical imaging setup is 2.5 $\mu$m and the error in the estimation of the peak magnetic field is $\pm 7$ MG.}}
\end{figure}

The magnetic fields observed at the rear surface are believed to be generated in conjunction with the electrostatic sheath field, which is set up when the fast electron beam generated at the front surface impinges on the rear surface \cite{DaviesPRE1999}.  As the fast electron beam has a finite transverse extent, so will the sheath field at early times.  This will lead to a significant net $\nabla \times {\bf E}$, which will generate a large magnetic field ($\p\tb B/\p t$), an analysis of which is given in Ref. \cite{RidgersPRE2011}. A simplistic order-of-magnitude estimate of the magnitude of these fields identifies their mechanism of generation, as follows. Typically, $B \approx E_{sheath}/c$, and as $E_{sheath}$ is of the order of a few TV/m \cite{MoraPRL2003}, we see that $B>10$ MG, which agrees with the magnitude of the magnetic fields observed in our experiments. The structure of the magnetic fields is determined by the transverse structure of the net fast electron beam \cite{GizziPRSTAB2011}.  A fast electron beam with a smooth profile about a well-defined center should produce an annular pattern (of azimuthal magnetic field), which peaks at some distance from the center, where there is a null in the field.  This would account for the patterns seen on the plastic targets, and suggest that there is some degree of filamentation in the aluminum targets.  As this would appear to be somewhat at odds with previous studies \cite{FuchsPRL2003, ManclossiPRL2006}, detailed fast electron transport calculations were carried out to see if these conclusions could be justified further.

\begin{figure}[b]
\includegraphics[width=\columnwidth]{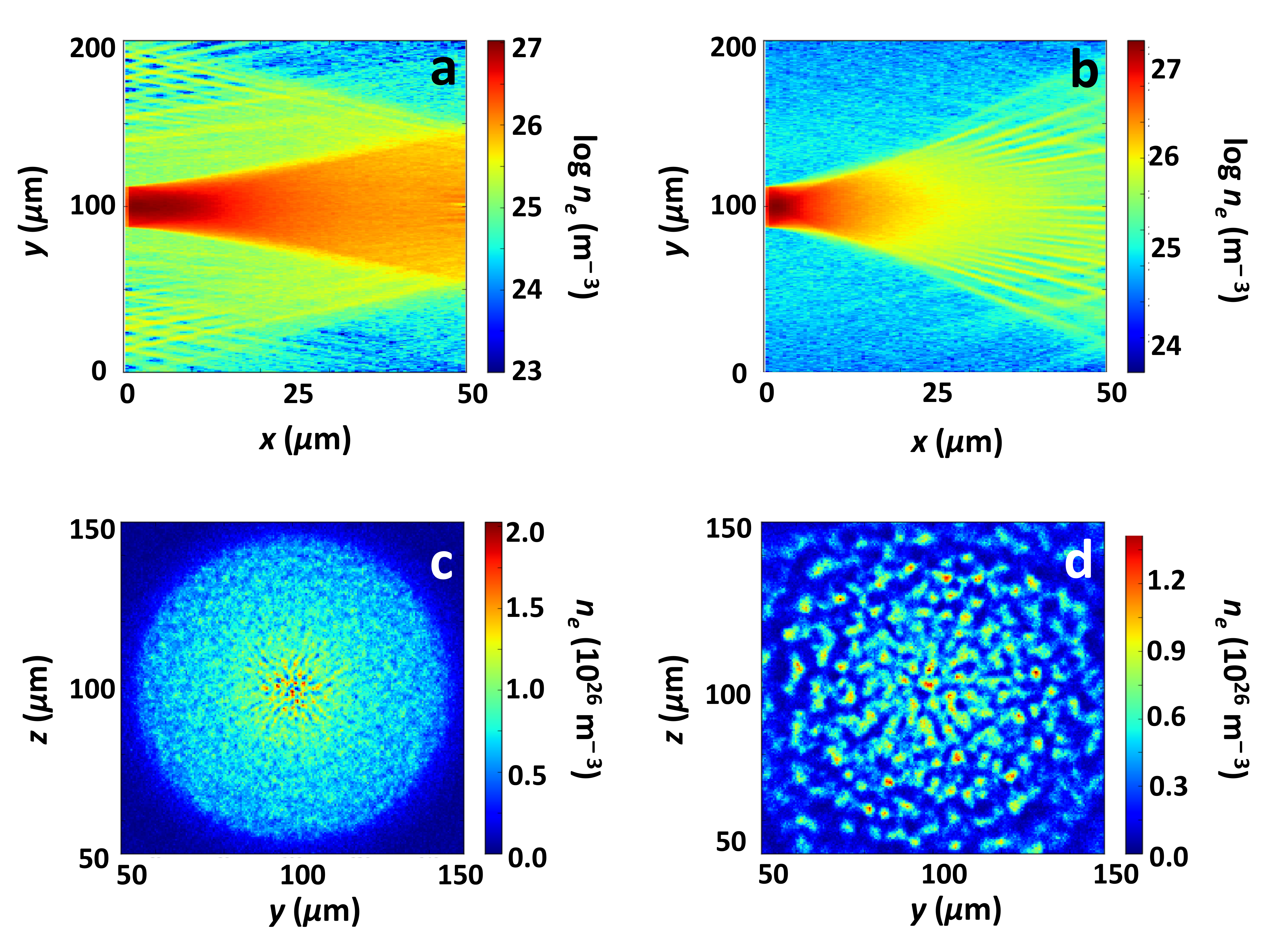}
\caption{\label{fig:sim_fig}\footnotesize{Results of the 3D hybrid simulation of fast electron transport at 0.6 ps in plastic (left pane) and aluminum (right pane).  Top figures [(a) and (b)] show longitudinal and bottom figures [(c) and (d)] show transverse profiles.}}
\end{figure}


\begin{figure*}[t]
\includegraphics[width=\textwidth]{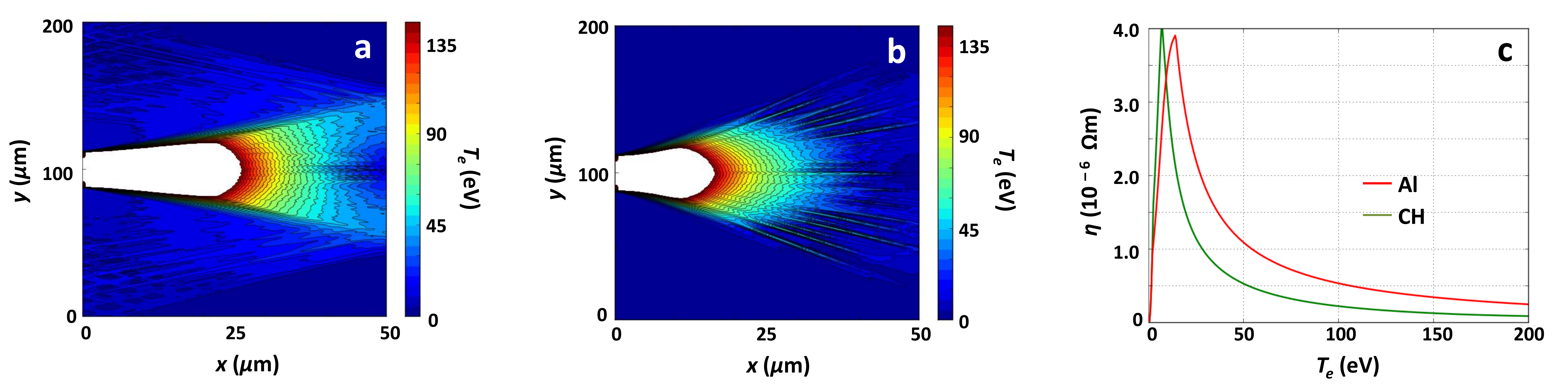}
\caption{\label{fig:resistivity_fig}\footnotesize{Temperature contours in the $x-y$ plane along the longitudinal direction in (a) plastic and (b) aluminum, 0.6 ps after the interaction. (c) Resistivities of plastic (green) and aluminum (red) as a function of bulk temperature. The saturated region indicates temperatures greater than 150 eV.}}
\end{figure*}


Fast electron transport calculations were carried out using {\sc zephyros}, a 3D hybrid code \cite{DaviesPRE2002,KarPRL2008}.  The simulations were performed using a 100$\times$400$\times$400 box with a cell size of 0.5$\times$0.5$\times$0.5~$\mu$m.  Fast electrons were injected from a region in the center of the $x=0$ plane so as to model laser irradiation at $1\times 10^{20}$ W/cm$^2$. Detailed simulations were carried out up to 0.8 ps, and this early period was found to be critical to whether filamentation was observed or not. A laser-to-fast-electron conversion efficiency of 30\% was assumed.  The transverse `laser spot' profile was a Gaussian function with an FWHM of 4 $\mu$m.  The fast electron energy distribution used was an exponential distribution ($\propto \exp(-\epsilon/\bar{\epsilon})$) with the mean energy, $\bar{\epsilon}$, determined by the ponderomotive scaling.  The angular distribution of the fast electrons was uniform over a solid angle subtended by a half-angle of 70$^\circ$ \cite{GreenPRL2008, OvchinnikovPRL2013}.  The background materials used were plastic (mylar) and aluminum.  In both simulations, the background temperature was initially 1~eV.  The resistivity curves were determined using the Lee-More model \cite{LeeMore1984}.  The $x$-boundaries of the simulation box were reflective to allow refluxing, but the transverse boundaries were open.

The results of the simulations are shown in Fig. \ref{fig:sim_fig}, where Figs. 4a and 4b give the longitudinal snapshots of the electron distribution at 0.6 ps after the interaction, as the electrons propagate through the thickness of the targets. It is clear that, although the electrons diverge as they pass through the material, the distribution remains beam-like in plastic, as shown in Fig. 4a. In aluminum, however, the electron beam seems to get fragmented after propagating beyond 30 $\mu$m (Fig. 4(b)). The transverse fast electron density profiles at the rear surface (Fig. 4(c) and 4(d)) illustrate this clearly. While the fast electron beam in the plastic target is relatively smooth, with minor levels of filamentation, as shown in Fig. 4c, the fast electron beam in the aluminum target exhibits very distinct small-scale filamentation  (Fig. 4(d)).  

Filamentation of electron currents in these experimental conditions is very likely to result from the resistive filamentation mechanism \cite{GremilletPOP2002}, which is most significant in such highly collisional regimes, where the two-stream instability is strongly suppressed \cite{RobinsonNucFus2013}.  The filamentation probability thus depends on the local resistivity profile inside the material. At high temperatures, the resistivity profiles of materials can be very different from those at room temperature.  For instance, aluminum has a much lower resistivity compared to plastic at room temperature but as the temperature increases, the resistivity profiles are significantly different. 

Figures \ref{fig:resistivity_fig}(a) and (b) show the background electron temperature contour plots in (a) plastic and (b) aluminum in the $x-y$ plane at 0.6 ps after the interaction. The saturated area  closer to the interaction reaches temperatures greater than 150 eV. The electron transport simulations presented in Fig. \ref{fig:sim_fig} show that  filamentation occurs at a depth of 30-40 $\mu$m. For both plastic and aluminum, the temperatures get to 40-75 eV in this region, depending on the distance from the interaction point in the $x$-direction. Figure \ref{fig:resistivity_fig}(c) shows the resistivity of plastic (green) and aluminum (red) as a function of temperature, using the Lee-More resistivity model \cite{LeeMore1984}. Although plastic is more resistive than aluminum at lower temperatures, aluminum becomes more resistive above 20 eV.  The fast electron beams would, therefore, be susceptible to filamentation beyond 30 $\mu$m in aluminum where the temperature is well above 20 eV; this is consistent with our experimental observations. 

In conclusion, we have explored the magnetic fields at the rear of solid targets, generated by fast electrons originating from the intense laser interaction at the target front surface. The optical polarimetry we employed is a sensitive technique that allows us to resolve the dynamics of electron propagation  with high  spatial and temporal resolution. As a result, we see, for the first time, the electron propagation through a conductor being  subjected to resistive filamentation in a regime where it remains approximately beam-like in an insulator. Specifically, our results identify an interaction regime in terms of local temperature where a metal like aluminum is clearly unstable to the resistive filamentation instability. In principle, such regimes could exist for any conductor and the current simplistic understanding that metals are generically more efficient carriers of fast electron currents is not universally applicable. For instance, the fast electron distribution not only depends on the initial conductivities of the materials but also on how the conductivity changes with temperature. In fact, the local temperature or resistivity inside the solid is inherently transient and is expected to be a complex function of the distance from the interaction point, local lattice configurations and laser parameters like intensity, pulsewidth and contrast.  For example, recent works report the dependence of fast electron distribution in solids on laser contrast  \cite{Scott_PRL_2012,OvchinnikovPRL2013} and local lattice order  \cite{McKennaPRL2011}. It is therefore essential to have a diagnostic that can unravel the complex dynamics of electron propagation through solids in order to optimise it. This is of critical importance in  developing novel sources for fast ions and engineering innovative techniques for long-range energy transport \cite{Gourab_PRL2012}. The experimental snapshots at fixed times presented here highlight the complexity in the phenomenon and suggest that it is highly transient in nature, yet amenable to accurate and detailed measurement. This measurement technique would enable us to extend these studies to obtain a full spatio-temporal understanding and a potential control of the fast electron transport process that is so central in laser-plasma research; further investigations to that end are under way.

The authors acknowledge the excellent experimental support provided by the Vulcan/Experimental Science staff at CLF. G.R.K acknowledges financial support from a J. C. Bose grant (DST, Govt. of India) and G.C. and P.K.S acknowledge support from the ``Strong Field Science" program (11P-1401). EPSRC support for the Fusion Doctoral Training Network is also gratefully acknowledged. P.K., L.L. and L.A.G. acknowledge financial support from MiUR project PRIN-2009FCC9MS.


\begin{thebibliography}{100}
\bibitem{ellipticity} Briefly, the phase difference between the ordinary $O$ wave and the extra-ordinary $X$ wave of the external probe, reflecting from the critical surface and the $X$ wave cutoff respectively, induces an ellipticity in the probe, which can be expressed in terms of the differences between the refractive indices of the $O$ and $X$ waves in accordance with the Appleton-Hartree formula. Since the refractive index of the $X$ wave depends on the ambient magnetic field via the cyclotron frequency, the ellipticity induced in the probe can be uniquely mapped on to the magnetic field experienced by the probe.
\end{thebibliography}
\end{document}